\documentclass[12pt]{article}

\usepackage{amssymb,amsmath}
\usepackage{cite}

\usepackage{graphicx}

\newcommand{\be}{\begin{equation}}
\newcommand{\ee}{\end{equation}}
\newcommand{\Dlt}{\Delta}
\newcommand{\dlt}{\delta}
\newcommand{\prt}{\partial}
\newcommand{\br}{{\bf r}}
\newcommand{\bj}{{\bf j}}

\newcommand{\bt}{\beta}
\newcommand{\vp}{\varphi}
\newcommand{\ep}{\varepsilon}
\newcommand{\al}{\alpha}

\newcommand{\sgm}{\sigma}

\newcommand{\om}{\omega}
\newcommand{\Om}{\Omega}
\newcommand{\Gm}{\Gamma}
\newcommand{\dgr}{\dagger}
\newcommand{\lbd}{\lambda}
\newcommand{\Lbd}{\Lambda}

\newcommand{\lgl}{\langle}
\newcommand{\rgl}{\rangle}

\begin{document}

\begin{center}

{\Large{\bf From Coherent Modes to Turbulence and Granulation 
of Trapped Gases } \\ [5mm]

V.S. Bagnato$^{1}$ and V.I. Yukalov$^{1,2}$} \\ [3mm]

{\it
$^1$Instituto de Fisica de S\~{a}o Calros, Universidade de S\~{a}o Paulo, \\
CP 369, 13560-970 S\~{a}o Carlos, S\~{a}o Paulo, Brazil

$^2$Bogolubov Laboratory of Theoretical Physics, \\
Joint Institute for Nuclear Research, Dubna 141980, Russia} \\ [3mm]

\end{center}

\vskip 3cm

\begin{abstract}
The process of exciting the gas of trapped bosons from an equilibrium
initial state to strongly nonequilibrium states is described as a
procedure of symmetry restoration caused by external perturbations.
Initially, the trapped gas is cooled down to such low temperatures, when
practically all atoms are in Bose-Einstein condensed state, which implies
the broken global gauge symmetry. Excitations are realized either by
imposing external alternating fields, modulating the trapping potential
and shaking the cloud of trapped atoms, or it can be done by varying atomic
interactions by means of Feshbach resonance techniques. Gradually
increasing the amount of energy pumped into the system, which is realized
either by strengthening the modulation amplitude or by increasing the
excitation time, produces a series of nonequilibrium states, with the
growing fraction of atoms for which the gauge symmetry is restored.
In this way, the initial equilibrium system, with the broken gauge
symmetry and all atoms condensed, can be excited to the state, where all
atoms are in the normal state, with completely restored gauge symmetry.
In this process, the system, starting from the regular superfluid state,
passes through the states of vortex superfluid, turbulent superfluid,
heterophase granular fluid, to the state of normal chaotic fluid in
turbulent regime. Both theoretical and experimental studies are presented.

\end{abstract}

\section{Introduction}

Different thermodynamic phases are usually characterized by different
symmetries. At the point of a phase transition, there occurs the change
of system symmetry \cite{Landau_1,Yukalov_2,Sornette_3}. The observation
of phase transitions can be done by slowly varying the system parameters,
e.g., temperature, pressure, density, or some stationary external fields,
so that the system practically always is in equilibrium

Another possibility of observing phase transitions is to prepare a system
in a nonequilibrium phase under the values of parameters favoring a
different phase. Then the system, starting from one phase with a given
symmetry, relaxes to the equilibrium phase with another symmetry,
dynamically passing through the phase-transition line \cite{Polkovnikov_4}.

In the present paper, we suggest and study the third way of realizing
phase transitions accompanied by symmetry changes. This way is opposite
to the relaxation procedure. We can start from an equilibrium phase, with
one type of symmetry, and then pump into the system energy by means of
external alternating fields, so that to transfer the system into another
state, with another symmetry type. We illustrate this idea by considering
the system of trapped bosons. This system can be cooled down to very low
temperatures below the Bose-Einstein condensation point, when all atoms
pile down to the condensed state. The properties of these condensed atoms
have been intensively studied both theoretically and experimentally, as
can be inferred from the books \cite{Pitaevskii_5,Lieb_6,Letokhov_7,
Pethik_8} and reviews \cite{Courteille_9,Andersen_10,Yukalov_11,Bongs_12,
Yukalov_13,Posazhennikova_14,Yukalov_15,Proukakis_16,Yurovsky_17,Yukalov_18,
Fetter_19,Yukalov_20}.

The Bose-condensed state is characterized by the global gauge symmetry
breaking. Moreover, the latter is the necessary and sufficient condition
for Bose-Einstein condensation \cite{Lieb_6,Yukalov_15}. Acting on the
system of trapped atoms by external alternating fields increases the system
energy, which is similar to increasing the system temperature. The energy,
pumped into the system, destroys the condensate, transferring atoms into
uncondensed states. When the injected energy is very large, one should
expect that the state can be reached where all condensate has been depleted,
and the whole system is in the normal phase, with the restored gauge symmetry.
This latter state will, of course, be nonequilibrium, being reached by
subjecting the system with time-dependent alternating fields. The
investigation of such a procedure of nonequilibrium transitions, going
through several stages, is the aim of the present paper. We shall describe
both theoretical as well as experimental peculiarities of this method. The
main part of the paper summarizes the results of previous publications, while 
some experimental results, related to the granular state, are new.

\section{Broken Gauge Symmetry}

We consider a system of spinless bosons characterized by the field operators
$\hat{\psi}({\bf r},t)$ satisfying Bose commutation relations. Here ${\bf r}$ is
spatial variable and $t$ is time. In the equations below, for the compactness
of notation, we often omit the time variable, assuming it but writing the field
operator as $\hat{\psi}({\bf r})$, when this does not lead to confusion. We keep
in mind dilute Bose gas confined in a trap modelled by an external trapping 
potential $U = U({\bf r},t)$. Atomic interactions are described by the local 
potential
\be
\label{1}
 \Phi(\br) = \Phi_0 \dlt(\br) \; , \qquad
\Phi_0 \equiv 4\pi \; \frac{a_s}{m} \;  ,
\ee
where $a_s$ is scattering length and $m$, atomic mass. The scattering length,
for concreteness, is assumed to be positive. Generally, it could be negative, 
but then the number of atoms should be such that to avoid the collapse occurring
for atoms with attractive interactions.   

Here and in what follows, we shall use in the majority of equations, the system 
of units with the Planck and Boltzmann constants set to one $(\hbar = 1, k_B = 1)$. 

The external potential consists of two terms,
\be
\label{2}
 U(\br,t) = U(\br) + V(\br,t) \;  ,
\ee
the first term being the trapping potential and the second term describing
additional modulation potential pumping energy into the trap.

The energy operator is given by the standard Hamiltonian
\be
\label{3}
 \hat H = \int \hat \psi(\br) \left ( - \; \frac{\nabla^2}{2m} + U
\right ) \hat\psi(\br) \; d\br +
\frac{\Phi_0}{2} \int \hat\psi^\dgr(\br) \hat\psi^\dgr(\br)
\hat\psi(\br)\hat\psi(\br) \; d\br \;  ,
\ee
where $U$ is the total external potential (2). In the presence of Bose-Einstein
condensate, the system global gauge symmetry is necessarily broken
\cite{Lieb_6,Yukalov_15}. The most convenient way of breaking the gauge symmetry
is by employing the Bogolubov shift \cite{Bogolubov_21,Bogolubov_22} of the
field operator:
\be
\label{4}
 \hat\psi(\br) = \eta(\br) + \psi_1(\br) \;  ,
\ee
in which $\eta({\bf r})$ is the condensate wave function normalized to the
number of condensed atoms
\be
\label{5}
 N_0 = \int | \eta(\br) |^2 d\br \;  ,
\ee
and $\psi_1({\bf r})$ is the operator of uncondensed atoms defining their number
\be
\label{6}
 N_1 = \lgl \hat N_1 \rgl \; , \qquad
\hat N_1 \equiv \int \psi_1^\dgr(\br) \psi_1(\br) \; d\br \;  ,
\ee
with the angle brackets implying statistical averaging. By this definition, the
field operator of uncondensed atoms satisfies the Bose commutation relations.

The condensate function and the field operator of uncondensed atoms characterize
different degrees of freedom, orthogonal to each other,
$$
 \int \eta^*(\br) \psi_1(\br) \; d\br = 0 \;  .
$$
The condensate function plays the role of the system order parameter, such that
\be
\label{7}
  \lgl \hat{\psi}(\br) \rgl = \eta(\br) \; , \qquad
\lgl \psi_1(\br) \rgl = 0 \;  .
\ee
This definition can also be written in the form of the statistical average
\be
\label{8}
\lgl \hat\Lbd   \rgl = 0
\ee
of the operator
\be
\label{9}
 \hat\Lbd \equiv \int \left [ \lbd(\br) \psi_1^\dgr(\br) +
\lbd^*(\br) \psi_1(\br) \right ] \; d\br \;  ,
\ee
in which $\lambda({\bf r})$ is a Lagrange multiplier guaranteeing the validity
of condition (7).

The correct description of any statistical system presupposes the use of the
representative ensemble uniquely defining the system
\cite{Yukalov_20,Yukalov_23,Yukalov_24,Yukalov_25}. This requires to take into
account all imposed constraints that, in the present case, are given by
Eqs. (5), (6), and (8). In turn, taking account of these constraints makes it
necessary to introduce the grand Hamiltonian
\be
\label{10}
 H = \hat H - \mu_0 N_0 - \mu_1 \hat N_1 - \hat\Lbd \;  ,
\ee
with the Lagrange multipliers $\mu_0$ and $\mu_1$. Only employing this grand
Hamiltonian allows one to correctly describe the Bose-condensed system. When
one uses an ensemble that is not representative, that is, when not all  
constraints are taken into account, this leads to various inconsistencies in 
thermodynamic and dynamic characteristics, such as the arising gap in the 
spectrum of elementary excitations and instability of the system.

\section{Nonequilibrium Bose System}

In the presence of an external time-dependent potential, we have to study a
nonequilibrium Bose system. The equations of motion for the system variables
can be written through the variational derivatives, which is equivalent to
the Heisenberg equations of motion \cite{Yukalov_20,Yukalov_26}. The
condensate function satisfies the equation
\be
\label{11}
i\; \frac{\prt}{\prt t} \; \eta(\br,t) =
\left \lgl \frac{\dlt H}{\dlt\eta^*(\br,t)} \right \rgl \;  .
\ee
While for the field operator of uncondensed atoms, one has
\be
\label{12}
 i\; \frac{\prt}{\prt t} \; \psi_1(\br,t) =
 \frac{\dlt H}{\dlt\psi_1^\dgr(\br,t)}  \;   .
\ee

To represent the resulting evolution equations in a convenient form, let us
introduce several notations. The condensate density is
\be
\label{13}
 \rho_0(\br) \equiv | \eta(\br) |^2 \;  .
\ee
The density of uncondensed atoms reads as
\be
\label{14}
 \rho_1(\br) \equiv \lgl \psi_1^\dgr(\br) \psi_1(\br) \rgl \;  .
\ee
When the gauge symmetry is broken, there appear the anomalous averages,
such as the pair anomalous average
\be
\label{15}
\sgm_1(\br) \equiv \lgl \psi_1(\br) \psi_1(\br) \rgl
\ee
and the triple anomalous averages
\be
\label{16}
 \xi(\br) \equiv \lgl \psi_1^\dgr(\br) \psi_1(\br) \psi_1(\br) \rgl \; ,
\qquad
\xi_1(\br) \equiv \lgl \psi_1(\br) \psi_1(\br) \psi_1(\br) \rgl \;  .
\ee
The total atomic density is the sum
\be
\label{17}
\rho(\br) = \rho_0(\br) + \rho_1(\br) \; .
\ee

Equation (11) yields the equation for the condensate function
\be
\label{18}
i\; \frac{\prt}{\prt t} \; \eta(\br) =
\left ( -\; \frac{\nabla^2}{2m} + U - \mu_0 \right ) \eta(\br) +
\Phi_0 \left [ \rho_0(\br) \eta(\br) + 2\rho_1(\br) \eta(\br) +
\sgm_1(\br) \eta^*(\br) + \xi(\br ) \right ] \;   .
\ee
And using Eq. (12), we find the continuity equation for the density of
uncondensed atoms,
\be
\label{19}
 \frac{\prt}{\prt t} \; \rho_1(\br) +
{\bf\nabla}\cdot\bj_1(\br) = - \Gm(\br) \;  ,
\ee
with the atomic current
\be
\label{20}
\bj_1(\br) \equiv -\; \frac{i}{2m}
\left \lgl \psi_1^\dgr(\br){\bf\nabla}\psi_1(\br) -
\left [ {\bf\nabla}\psi_1^\dgr(\br) \right ] \psi_1(\br) \right \rgl
\ee
and the source term given by the expression
\be
\label{21}
 \Gm(\br) = i \Phi_0 \left [ \Xi^*(\br) - \Xi(\br) \right ] \;  ,
\ee
in which
\be
\label{22}
 \Xi(\br) \equiv
\eta^*(\br) \left [ \eta^*(\br) \sgm_1(\br) + \xi(\br) \right ] \;  .
\ee

In addition, it is necessary to consider the equations for the anomalous
averages. Writing down the equation for the pair average (15), we can use
the Hartree-Fock-Bogolubov approximation for the four-operator correlator
\be
\label{23}
 \lgl \psi_1^\dgr(\br) \psi_1(\br) \psi_1(\br) \psi_1(\br) \rgl =
3\rho_1(\br) \sgm_1(\br) \;  .
\ee
Also, we define the anomalous kinetic term
$$
K(\br) \equiv -\;
\frac{1}{2} \left \lgl \frac{\nabla^2\psi_1(\br)}{2m} \; \psi_1(\br) +
\psi_1(\br) \; \frac{\nabla^2\psi_1(\br)}{2m} \right \rgl \;  =
$$
\be
\label{24}
 = \;
\frac{1}{2m} \left \{ \left \lgl [ \nabla\psi_1(\br)^2  ]\right \rgl -\;
\frac{1}{2} \; \nabla^2\sgm_1(\br) \right \} \;  .
\ee
Then the evolution equation for the anomalous average (15) is
$$
i\; \frac{\prt}{\prt t} \; \sgm_1(\br) = 2 K(\br) +
2 ( U - \mu_1 ) \sgm_1(\br) \; +
$$
\be
\label{25}
+ \; 2\Phi_0 \left [ \eta^2(\br) \rho_1(\br) + 2\rho_0(\br) \sgm_1(\br) +
3\rho_1(\br) \sgm_1(\br) + 2 \eta(\br) \xi(\br) +
\eta^*(\br) \xi_1(\br) \right ] \;   .
\ee
Equations (18) to (25) describe the nonequilibrium system with Bose-Einstein
condensate \cite{Yukalov_20}.

\section{Topological Coherent Modes}

Strongly nonlinear time-dependent equations, such as the condensate-function
equation (18), can display different nonequilibrium solutions. One usually
considers a particular case of this equation corresponding to asymptotically
weak interactions, when one can neglect the terms containing $\rho_1$ and
$\sigma_1$. In that case, Eq. (18) reduces to the nonlinear Schr\"{o}dinger
equation, also called the Gross-Pitaevskii equation
\cite{Gross_27,Gross_28,Ginzburg_29,Gross_30,Pitaevskii_31}. Such a nonlinear
equation possesses a variety of soliton solutions \cite{Malomed_32,Kartashov_33}.
Here we shall consider a special class of nonequilibrium solutions that can
exist being supported by the action of external alternating fields.

First, let us define the set of stationary solutions to the condensate-function
equation (18). These solutions are obtained by considering the situation
without the time-dependent perturbation $V({\bf r},t)$ and substituting into
Eq. (18) the form
\be
\label{26}
 \eta_n(\br,t)  = \eta_n(\br) e^{-i\om_n t} \; ,
\ee
which results in the eigenvalue problem
$$
\left [ -\; \frac{\nabla^2}{2m} + U(\br) \right ] \eta_n(\br) \; +
$$
\be
\label{27}
 + \;
\Phi_0 \left [ | \eta_n(\br)|^2 \eta_n(\br) + 2 \rho_1(\br) \eta_n(\br)
+ \sgm_1(\br) \eta_n^*(\br) + \xi(\br) \right ] =
E_n \eta_n(\br) \;  ,
\ee
where $n$ is a multi-index labelling the eigenstates and
\be
\label{28}
 E_n \equiv \om_n + \mu_0 \;  .
\ee
The related stationary solutions for $\rho_1$ and $\sigma_1$ are assumed to
enter Eq. (27), or they are neglected in the simple case of the Gross-Pitaevskii
equation. The lowest eigenvalue $E_n$ corresponds to the equilibrium case, when
\be
\label{29}
 \mu_0 = \min_n E_n \qquad (\min_n \om_n = 0 ) \;  .
\ee

The solutions to Eq. (27) are termed {\it coherent topological modes}. They
are coherent, since the condensate function corresponds to the coherent state,
in agreement with the general definition of such states \cite{Klauder_34}. And
they are called topological because the solutions with different indices $n$
possess different spatial topology, having different number of zeroes.
Respectively, the related densities of condensed atoms $|\eta_n({\bf r})|^2$,
with differing indices $n$, have different spatial shapes. The coherent
topological modes for the Gross-Pitaevskii equation were introduced in
Ref. \cite{Yukalov_35}; and their properties were studied in many articles
\cite{Yukalov_36,Ostrovskaya_37,Feder_38,Yukalov_39,Kivshar_40,Agosta_41,
Yukalov_42,Yukalov_43,Damski_44,Agosta_45,Yukalov_46,Proukakis_47,Yukalov_48,
Yukalov_49,Adhikari_50,Adhikari_51,Muruganandam_52,Yukalov_53,Yukalov_54,
Adhikari_55,Filho_56,Ramos_57,Ramos_58,Ramos_59,Ramos_60,Yukalov_61}.
A dipole topological mode was excited in experiment \cite{Williams_62}.
The general case of Eq. (27) has also been considered
\cite{Yukalov_20,Yukalov_63}.

As an illustration of typical solutions, representing such coherent modes,
let us consider the case of zero temperature and weak atomic interactions,
when the Gross-Pitaevskii equation is applicable. The atoms are trapped in 
a harmonic cylindrical trapping potential. The corresponding solution can 
be represented \cite{Courteille_9,Yukalov_35,Yukalov_36,Yukalov_43,Yukalov_46} 
in the form
$$
\psi_{nmj}(r,\vp,z) = \left [ \frac{2n! u^{|m|+1}}{(n+|m|)!} \right ]^{1/2}
r^{|m|} \exp \left ( - \; \frac{u}{2}\; r^2 \right ) L_n^{|m|} \left (ur^2
\right ) \times
$$
$$
\times \; \frac{e^{im\vp}}{\sqrt{2\pi}} \; \left ( \frac{v}{\pi}\right )^{1/4}
 \frac{1}{\sqrt{2^j j!} } \; \exp \left ( -\; \frac{v}{2} \; z^2 \right )
H_j(\sqrt{v}\; z) \;   ,
$$
in which $L_n^m$ is a Laguerre polynomial, $H_j$, a Hermite polynomial,
$n=0,1,2,\ldots$ is the radial quantum number, $m= 0,\pm 1, \pm 2, \ldots$
is the azimuthal quantum number, and $j = 0,1,2,\ldots$ is the axial quantum
number. The variables $r, \varphi, z$ are cylindrical coordinates. And the 
quantities $u, v$ are the so-called control functions, depending on all 
system parameters and defined so that to guarantee the convergence of 
optimized perturbation theory 
\cite{Yukalov_p64,Yukalov_p65,Yukalov_p66,Yukalov_p67}. As is clear, the
solutions with nonzero azimuthal quantum number $m$ correspond to vortices. 

When there is no external perturbation, the system tends to its equilibrium
state corresponding to the lowest energy level (29). But if the system is
subject to an external time-dependent perturbation, then we have to consider
the evolution equation (18). It is admissible to look for the solution of
this equation in the form of the expansion over the coherent modes:
\be
\label{30}
 \eta(\br,t) = \sum_n c_n(t) \eta_n(\br) e^{-i\om_n t} \;  .
\ee

Defining the number of condensed atoms at the initial time,
\be
\label{31}
 N_0 \equiv \int | \eta(\br,0) |^2 d\br \;  ,
\ee
we use the notation
\be
\label{32}
\eta_n(\br) = \sqrt{N_0} \; \vp_n(\br) \;   ,
\ee
introducing the functions $\varphi_n$ normalized to one:
$$
 \int | \vp_n(\br) |^2 d\br = 1\;  .
$$
Note that these functions $\varphi_n$, being defined by a nonlinear equation,
are not necessarily orthogonal.

We impose, in addition to the stationary trapping potential $U({\bf r})$, the 
external potential modulating the trapping potential in the form
\be
\label{33}
  V(\br,t) = V_1(\br) \cos(\om t) +
V_2(\br) \sin(\om t) \; ,
\ee
with the total potential given by Eq. (2). Also, we require that this 
alternating potential be in resonance with one of the transition frequencies 
$\omega_n$, so that the resonance condition
\be
\label{34}
 \left | \frac{\Dlt\om}{\om} \right | \ll 1 \qquad
( \Dlt\om \equiv \om - \om_n)
\ee
be valid for the fixed $n$. Substituting expansion (30) into Eq. (18) and
employing the averaging techniques
\cite{Bogolubov_64,Yukalov_65,Yukalov_66}, we come to the equations
$$
i\; \frac{dc_0}{dt} = \al_{0n} | c_n|^2 c_0 +
\frac{1}{2} \; \bt_{0n} c_n e^{i\Dlt\om \cdot t} \; ,
$$
\be
\label{35}
i\; \frac{dc_n}{dt} = \al_{n0} | c_0|^2 c_n +
\frac{1}{2} \; \bt^*_{0n} c_0 e^{-i\Dlt\om \cdot t} \;  ,
\ee
in which
$$
 \al_{mn} \equiv \Phi_0 N_0 \int | \vp_{m}(\br) |^2 \left [
2 | \vp_n(\br) |^2 - |\vp_m(\br)|^2 \right ] \; d\br \;  ,
$$
$$
 \bt_{mn} \equiv \int  \vp^*_{m}(\br) \left [
V_1(\br) - i V_2(\br)  \right ]\vp_n(\br) \; d\br \; .
$$
Solving these equations gives us the fractional mode populations
\be
\label{36}
p_n(t) = | c_n(t) |^2 \; , \qquad
\sum_n p_n(t) = 1\;   .
\ee
It is worth noting that the mathematical structure of these equations is
the same as that of equations describing atomic motion in a double-well
potential. Therefore solutions to these equations exhibit many properties
that are analogous to the properties of solutions in the case of a
double-well potential. For instance, the effect of mode locking
\cite{Yukalov_35,Yukalov_46}, occurring for Eqs. (35), is mathematically
identical to the effect of self-trapping for the double well potential
\cite{Smerzi_67}.

Among other interesting effects, exhibited by the system with the generated
coherent topological modes, we can mention the interference patterns and
interference current \cite{Yukalov_42,Yukalov_43,Yukalov_46}, dynamical
phase transitions and critical phenomena
\cite{Yukalov_39,Yukalov_42,Yukalov_43,Yukalov_46}, chaotic motion under
the action of several alternating fields \cite{Yukalov_53,Yukalov_54},
atomic squeezing \cite{Yukalov_46,Yukalov_48,Yukalov_49}, Ramsey fringes
\cite{Ramos_57,Ramos_58,Ramos_59}, and entanglement production
\cite{Yukalov_ 68,Yukalov_69,Yukalov_70,Yukalov_71}.

The coherent topological modes can also be generated by modulating the
atomic scattering length by means of the Feshbach resonance techniques
\cite{Yukalov_20,Ramos_60,Yukalov_61}, so that the interaction strength be
varying in time as
\be
\label{37}
\Phi(t) = \Phi_0 + \Phi_1\cos(\om t) + \Phi_2 \sin(\om t) \;   ,
\ee
provided that the alternating frequency $\omega$ is tuned close to one of
the transition frequencies $\omega_n$.

In the case of resonance $\omega = \omega_n$, coherent modes can be
generated by an external modulation of rather weak strength. But increasing
the amplitude of the pumping field simplifies this generation, making the
strict resonance $\omega = \omega_n$ not necessary
\cite{Yukalov_53,Yukalov_54}. Then several other conditions come into play
allowing for the mode generation. Thus, the modes can be created when the
external frequency is close to the condition of {\it harmonic generation}
\be
\label{38}
 k \om = \om_n \qquad (k = 1,2, \ldots ) \;  .
\ee
If there are two alternating fields, with the frequencies $\omega_1$ and
$\omega_2$, then the modes can be produced \cite{Yukalov_53,Yukalov_54}
by {\it parametric conversion}, when the frequencies satisfy 
(at least approximately) the relation
\be
\label{39}
 \om_1 \pm \om_2 = \om_n \;  .
\ee
This effect is similar to parametric resonance \cite{Baizakov_72}. 

Generally, for several alternating fields, with frequencies $\omega_i$,
the condition of the {\it generalized resonance}
\be
\label{40}
\sum_i k_i \om_i = \om_n \qquad
(k_i = \pm 1, \pm 2, \ldots )
\ee
is sufficient for generating coherent modes.

In this way, increasing the amplitude of the pumping field produces in the
trapped Bose gas a variety of different topological coherent modes. The same
multiple mode creation happens when the action of the alternating perturbing
potential lasts sufficiently long, during the time after which the effect
of power broadening comes into play
\cite{Yukalov_35,Yukalov_46,Yukalov_54,Yukalov_63}.

\section{Creation of Trapped Vortices}

One type of the coherent topological modes is of special interest. These
are the quantum vortices. Such vortices have been observed in superfluid
helium \cite{Yarmchuk_72} and in trapped Bose-Einstein condensate
\cite{Matthews_73,Madison_74,Rosenbusch_75}. In the dynamical picture,
the appearance of vortices is caused by a dynamical instability arising
in a nonequilibrium moving fluid
\cite{Yukalov_76,Dutton_77,Yukalov_78,Ruostekoski_79,Shomroni_80,Ma_81,
Ishiro_82,Simula_83}.

The first vortex appears, when the atomic cloud is rotated with the
frequency reaching the critical value $\omega_{vor}$. Let us consider a cloud
of Bose-condensed atoms in a cylindrical trap with a transverse,
$\omega_\perp$, and longitudinal, $\omega_z$, trap frequencies, and with the
aspect ratio
\be
\label{41}
 \al \equiv \frac{\om_z}{\om_\perp} =
\left ( \frac{l_\perp}{l_z} \right )^2 \; ,
\ee
in which the effective trap lengths are
$$
 l_\perp \equiv \frac{1}{\sqrt{m\om_\perp} } \; , \qquad
l_z \equiv \frac{1}{\sqrt{m\om_z} } \;   .
$$
The critical rotation frequency for this trap \cite{Pethik_8} can be written as
\be
\label{42}
  \om_{vor} =
\frac{5}{2mR_{TF}^2} \; \ln \left ( 0.7\; \frac{R_{TF}}{\xi} \right ) \; ,
\ee
where the notations are used for the Thomas-Fermi radius
\be
\label{43}
  R_{TF} = l_\perp \left ( \frac{15}{4\pi} \; \al g \right )^{1/5} \; ,
\ee
dimensionless coupling parameter
\be
\label{44}
  g \equiv 4\pi N \; \frac{a_s}{l_\perp} \; ,
\ee
and the healing length
\be
\label{45}
\xi \equiv \frac{1}{\sqrt{2m\rho(0)\Phi_0} } \;   .
\ee

The vortex with vorticity one is energetically more stable than the vortices
with higher vorticities. Because of this, the latter decay into several basic
vortices with vorticity one. Moreover, for large coupling parameter (44) the
basic vortex is the most stable among all coherent modes
\cite{Courteille_9,Yukalov_20}. This follows from the fact that the basic
vortex energy, that can be represented by Eq. (42), can be rewritten as
\be
\label{46}
 \om_{vor} = \frac{0.9\om_\perp}{(\al g)^{2/5} } \;
\ln( 0.8 \al g) \;  ,
\ee
which shows that this energy diminishes with $g$. While the energies of other
coherent modes increase with $g$ as
\be
\label{47}
 \om_n \propto (\al g) ^{2/5} \qquad (g \gg 1) \;  .
\ee
Increasing the velocity of rotation produces many basic vortices that form
arrays arranged into Abrikosov lattices \cite{Madison_74,Abo_84}.

However, if we modulate the trapping potential by alternating fields without
a fixed rotation axis, as is described above for generating coherent modes,
then we shall generate vortices and antivortices. Such a type of vortex creation
was demonstrated in experiments \cite{Henn_85,Seman_86}, where the harmonic
trapping potential
\be
\label{48}
 U(\br) = \frac{m}{2} \; \om_\perp^2 \left ( x^2 + y^2 \right ) +
\frac{m}{2} \; \om_z^2 z^2 \;  ,
\ee
with $\omega_\perp = 2 \pi \times 210$ Hz and $\omega_z = 2 \pi \times 23$ Hz,
was modulated with the alternating potential
\be
\label{49}
 V(\br,t) = \frac{m}{2} \; \Om_x^2(t) (x - x_0)^2 +
\frac{m}{2} \; \Om_y^2(t) (y' - y'_0 )^2 +
\frac{m}{2} \Om_z^2(t) (z' - z_0')^2 \; .
\ee
Here the oscillation centers are defined by the equation
$$
\left [ \begin{array}{c} y' - y_0' \\
                         z' - z_0' \end{array}  \right ]
= \left [ \begin{array}{cc} \sin\vartheta_0 & \cos\vartheta_0 \\
                            \cos\vartheta_0 & -\sin\vartheta_0 \end{array} \right ]
\left [ \begin{array}{c} y - y_0 \\
                         z - z_0 \end{array} \right ] \; ,
$$
and the oscillation frequencies are
\be
\label{50}
 \Om_\al(t) = \om_\al \dlt_\al [ 1 - 1 \cos(\om t) ] \;  ,
\ee
with $\alpha = x, y, z$, and $x_0, y_0, z_0, \vartheta_0, \delta_{\alpha}, \omega$
being fixed parameters \cite{Seman_87}.

\section{Trapped Turbulent Superfluid}

Strong rotation creates a vortex lattice \cite{Pethik_8}. But when the
trapped atomic cloud is subject to the action of an alternating
modulation potential without a fixed rotation axis and this pumping injects
into the system the amount of energy sufficient for creating many vortices
and antivortices, then the latter are randomly distributed inside the trap,
forming a chaotic tangle. Such a random tangle of vortices is associated with
turbulence, similar to the spatially tangled vortices in superfluid helium
\cite{Feynman_88}.

Turbulence is a phenomenon that has been studied for classical liquids for
many years \cite{Davidson_89}. Vortices in a classical fluid can be of
different vorticities, while the vortex circulation in a quantum fluid is
quantized, which distinguishes the classical turbulence from the {\it quantum
turbulence} \cite{Donnelly_90}.

One of important characteristics of turbulent motion is the mean kinetic energy
that can be represented as the integral
\be
\label{51}
K = \int_0^\infty E(k) \; dk
\ee
over the wave-number values $k$. In classical fluids, there exists a diapason
of wave numbers, called {\it inertial range}, where the spectrum $E(k)$, is
given by the Kolmogorov \cite{Kolmogorov_91,Kolmogorov_92} law for isotropic
turbulence
\be
\label{52}
 E(k) = C \ep^{2/3} k^{-5/3} \;  ,
\ee
with $C \approx 1.5$ and $\epsilon$ being energy transfer rate. The Kolmogorov
law is universal for classical fluids \cite{Sreenivasan_93}

Quantum turbulence was, first, studied for superfluid helium
\cite{Hall_94,Vinen_95}. It was found in experiments \cite{Maurer_96,Stalp_97}
that there also exists an inertial range of wave numbers, where the Kolmogorov
law (52) is valid, independently of temperature. In superfluids, the energy is
dissipated through the interaction of the normal and superfluid components
and, at low temperature, through vortex reconnection, Kelvin wave excitations,
and phonon emission \cite{Svistunov_98,Kobayashi_99}. More details can be found
in Refs. \cite{Barenghi_100,Vinen_101,Yukalov_102,Nowak_103}.

Numerical simulation of quantum turbulence in Bose-Einstein condensate is
usually done by solving the Gross-Pitaevskii equation. Atoms are assumed to be
trapped in a stationary trap and subject to the action of an external alternating
perturbation with more than one rotation axes. The kinetic energy, when all atoms
are condensed, is given by the integral
\be
\label{53}
 K = \int \eta^*(\br,t) \left ( -\; \frac{\nabla^2}{2m} \right )
\eta(\br,t) \; d\br \;  .
\ee
It was found \cite{Kobayashi_104,Tsubota_105} that there again exists an inertial
range, where the Kolmogorov law is applied. Thus, for atoms in a harmonic trap,
the inertial range is
\be
\label{54}
 \frac{2\pi}{R_{TF}} \; < k \; < \frac{2\pi}{\xi} \qquad (C \approx 0.25) \; ,
\ee
where $R_{TF}$ is the Thomas-Fermi radius and $\xi$, healing length. For atoms
in a box of linear size $L$, the inertial range is
\be
\label{55}
 \frac{2\pi}{L} \; < k \; < \frac{2\pi}{\xi} \qquad (C \approx 0.55) \;  .
\ee

Experimental generation of trapped quantum turbulence was realized
\cite{Henn_106,Seman_107,Shiozaki_108,Seman_109} for $^{87}$Rb Bose-Einstein
condensate. It was trapped in the harmonic potential (48) and subject to the
action of the alternating potential (49).

\section{Heterophase Granular Fluid}

If we continue pumping energy into the system, turbulence is getting stronger
and stronger. The core of each vortex can be treated as a nucleus of normal
(uncondensed) phase. Producing more and more vortices increases the amount
of the uncondensed component. What then happens, when the number of vortices
in the strongly turbulent liquid is so large that the amount of the
uncondensed fraction becomes comparable or greater than the fraction of
condensed atoms? The answer to this question cannot be done being based solely
on the Gross-Pitaevskii equation that describes only the condensed fraction.
To take into account both the condensed as well as uncondensed fractions, it
is necessary to consider the full evolution equations (18) to (25).

A simple way of understanding what happens in a strongly nonequilibrium system
under the action of a time-dependent perturbation is as follows. It is possible
to prove \cite{Yukalov_20,Yukalov_102,Yukalov_110} that the system with the
time-dependent perturbation can be mapped onto the system subject to the action
of a random spatial potential, provided that the modulation period is larger
than the local equilibration time. The behavior of the weakly interacting
Bose-condensed system in a weak spatially random potential has been studied in
several articles (see, e.g., \cite{Graham_111,Gaul_112}). A theory for Bose
systems with arbitrarily strong interactions and random potentials of arbitrary
strength has also been developed \cite{Yukalov_110,Yukalov_113,Yukalov_114}.

Using the analogy between the spatially random and temporally perturbed Bose
gas \cite{Yukalov_20,Yukalov_110} we can evaluate the {\it localization length}
defining the scale at which Bose gas can be condensed. This length for a trapped
Bose gas is
\be
\label{56}
 l_{loc} = \frac{1}{m^2E_{inj}^2 l_0^3} =
\left ( \frac{\om_0}{E_{inj}} \right )^2 l_0 \;  ,
\ee
where the effective trap size and effective frequency are
\be
\label{57}
 l_0 \equiv \left ( l_\perp^2 l_z \right )^{1/3} =
\frac{1}{\sqrt{m\om_0} } \; ,   \qquad
\om_0 \equiv \left ( \om_\perp^2 \om_z \right )^{1/3} =
\frac{1}{ml_0^2 } \;  ,
\ee
and the energy per atom, injected into the trap, can be evaluated as
\be
\label{58}
 E_{inj} \approx \frac{1}{N} \int \rho(\br,t) \left |
\frac{\prt V(\br,t)}{\prt t} \right | \; d\br dt \;  .
\ee
If the pumping potential is alternating, as is usual, with an amplitude $A$ and
frequency $\omega$, then the energy, injected in the time interval $[t,t^\prime]$,
is approximately
\be
\label{59}
E_{inj} \approx A \om(t - t') \;   .
\ee
This expression for the injected energy is certainly approximate, since a part
of the pumped energy is dispersed, but not transferred to atoms.

If the localization length (56) is larger or of order of the trap size, given
in Eq. (57), then all atoms in the trap are in the Bose-condensed state. But when
this length becomes shorter than the trap size, though yet larger than the mean
interatomic distance $a$, then the atomic cloud breaks into pieces. Then the
system consists of grains, composed of Bose-condensed phase, immersed into the
cloud, consisting of normal phase, without gauge symmetry breaking. The sizes of
the condensate grains are of order of the localization length. Thus, the condition
for the occurrence of this heterophase granular state is
\be
\label{60}
 a < l_{loc} < l_0 \; .
\ee
Such a heterophase state is similar to heterophase states arising in many
condensed-matter systems \cite{Yukalov_23,Yukalov_115} and that can happen in
optical lattices \cite{Yukalov_18,Yukalov_116}.

The state of the heterophase granular fluid has been observed in experiment 
\cite{Seman_109} with a cloud of strongly modulated $^{87}$Rb atoms.

\section{Normal Chaotic Fluid}

What happens, if we continue pumping energy into the trapped atomic cloud?
Again, following the analogy with other heterophase systems
\cite{Yukalov_18,Yukalov_23,Yukalov_115,Yukalov_116}, we should expect that
the fraction of the Bose-condensed phase, concentrated in the grains, will
diminish, and, finally, the whole system will be transferred into the normal
state, with the restored gauge symmetry. Being subject to strong external
perturbation, the system will, of course, be essentially nonequilibrium,
experiencing chaotic fluctuations. So, this will be a normal chaotic fluid,
with completely restored global gauge symmetry, without any remnants of
Bose-Einstein condensate,

The normal chaotic fluid could, probably, be characterized by the approach
called weak-turbulence theory, or wave-turbulence theory
\cite{Tsytovich_117,Dyachenko_118,Galtier_119,Berloff_120,Lvov_121,Zakharov_122}.
In this approach one assumes that turbulence in a weakly nonlinear system can
be represented by an ensemble of weakly interacting waves. However, the
nonlinearity in the system can be rather strong. And the normal chaotic state,
with no condensate, cannot be described by the Gross-Pitaevskii equation
appropriate only for pure condensate. More probably, the normal chaotic state
is just a strongly turbulent state of a normal fluid and could be described
as classical turbulence.

This state has not yet been reached in experiments \cite{Shiozaki_108,Seman_109}
and remains to be investigated.

\section{Amplitude-Time Phase Diagram}

The whole procedure of exciting the system of trapped atoms by applying an
external alternating perturbation potential passes through several stages.
We start with an almost completely Bose-condensed gas, where the global
gauge symmetry is broken. Very weak perturbation can do not more than to
produce elementary collective excitations that do not change the overall
system properties. This state can be called the {\it regular superfluid}.

When the energy, injected into the trap, becomes comparable with the energy
of a vortex, a single vortex is created. This happens when
$E_{inj} \sim \omega_{vor}$. With equality (59), this gives the relation
\be
\label{61}
A_{vor} \sim \frac{\om_{vor}}{\om(t-t_0)}
\ee
between the amplitude $A$ of the alternating perturbing potential and the
time $t$ of its action, describing the effective transition line of vortex
creation. Above this line, we have the state of {\it vortex superfluid}. Of
course, the transition from the regular superfluid to vortex superfluid is
not a sharp phase transition, but it is a crossover. However, the crossover
line (61) serves as an approximate separation line between these two
qualitatively different regimes. Similarly, the dividing lines between
other qualitatively different regions are also crossover lines.

Increasing the injected energy, pumped into the trap, by either a stronger
alternating field or by its longer action, leads to the generation of a
variety of coherent topological modes that decay into basic vortices and
antivortices. To create $N_{vor}$ vortices (and antivortices), it is
necessary to inject the energy $E_{inj} \sim N_{vor} \omega_{vor}$. When
the number of vortices becomes large, of order
\be
\label{62}
 N_{vor} \sim \frac{l_0}{\xi} \;  ,
\ee
they form a random tangle, which signifies the appearance of turbulent state.
Hence, the crossover line between the vortex superfluid and the
{\it turbulent superfluid} is given by
\be
\label{63}
 A_{tur} \sim \frac{l_0\om_{vor}}{\xi\om(t-t_1)} \;  .
\ee
The random vortex tangle is formed due to the property of the imposed
perturbing potential that does not prescribe a single rotation axis.

As soon as the injected energy reaches the value $E_{inj} \sim \omega_0$,
the condensate localization length (56) becomes of order of the trap size
$l_0$. As is explained in Sec. 7, in the region of the localization lengths
(60), the heterophase granular state arises. This {\it granular fluid}
consists of the grains of Bose-condensed gas immersed into the cloud of
normal fluid without gauge symmetry breaking. The corresponding crossover
line writes as
\be
\label{64}
 A_{het} \sim \frac{\om_0}{\om(t-t_2)} \;  .
\ee
The cloud of normal atoms, surrounding the Bose-condensed droplets, is
characterized by the restored gauge symmetry.

When the localization length (56) becomes as small as the mean interatomic
distance, no condensed droplets can be formed. That is, on the boundary,
where
$$
 l_{loc} \sim a \; , \qquad
E_{inj} \sim \om_0 \sqrt{\frac{l_0}{a} } \;  ,
$$
all condensate is completely destroyed. This defines the crossover line
\be
\label{65}
A_{nor} \sim \frac{\om_0}{\om(t-t_3)} \sqrt{\frac{l_0}{a} } \;
\ee
between the granular fluid and the normal fluid with no gauge symmetry
breaking. Since the latter is in a strongly nonequilibrium state with
chaotic motion, it can be termed {\it chaotic fluid}. This regime,
presumably, can be characterized by classical turbulent state.

Summarizing the sequence of these crossover transitions, we have:
$$
 \begin{array} {rl}
0 < A < A_{vor} & ~~(regular \; superfluid) \\
A_{vor} < A < A_{tur} & ~~(vortex \; superfluid) \\
A_{tur} < A < A_{het} & ~~(turbulent \; superfluid) \\
A_{het} < A < A_{nor} & ~~(granular \; fluid) \\
A > A_{nor} & ~~(chaotic \; fluid) \end{array} \; .
$$
The first four of these regimes have been observed in experiments, as
described above. The last state of chaotic fluid has not yet been
reached for trapped atoms.

\section{Experiments with Strongly Nonequilibrium Trapped Bose gas}

While classical turbulence can be observed quite easily with visualization
techniques, for traditional superfluids that is not the case. The high density
in superfluid liquid-He makes the vortex line core of order of atomic scale
dimensions, and therefore, turning the visualization techniques hard to be
applied. On the other hand, in trapped atomic superfluids the low density
makes possible the observation of vortex arrangement with unaided eye. We
therefore use the observations of irregular arrangement of vortices as
one of the macroscopic indications of Quantum Turbulence (QT). After the regime
of QT is reached, the studies of many aspects, revealing the similarities and
differences with the classical counterpart, become of great interest.

The first important aspect on the experimental observation of QT is the
production of vortex lines. The standard way of producing quantized vortices
in a trapped condensate is by stirring \cite{Madisona_47, Ram_87}. Laser beams
or rotation of an asymmetric trapping potential are the alternatives to achieve 
a rotating cloud of atoms. In these cases, the nucleation of vortices takes 
place in a specific direction (along the rotation axis), and therefore the final
result is a lattice of vortices instead of a tangle configuration. To achieve
a tangle configuration, we have developed a new technique \cite{Henn_85}, where
a combination of oscillations in the cloud results in the nucleation of vortices
in many directions, which is a necessary ingredient for the final production of
a tangle configuration of vortices.

\begin{figure}[!h]
\centering
\includegraphics[scale=0.25]{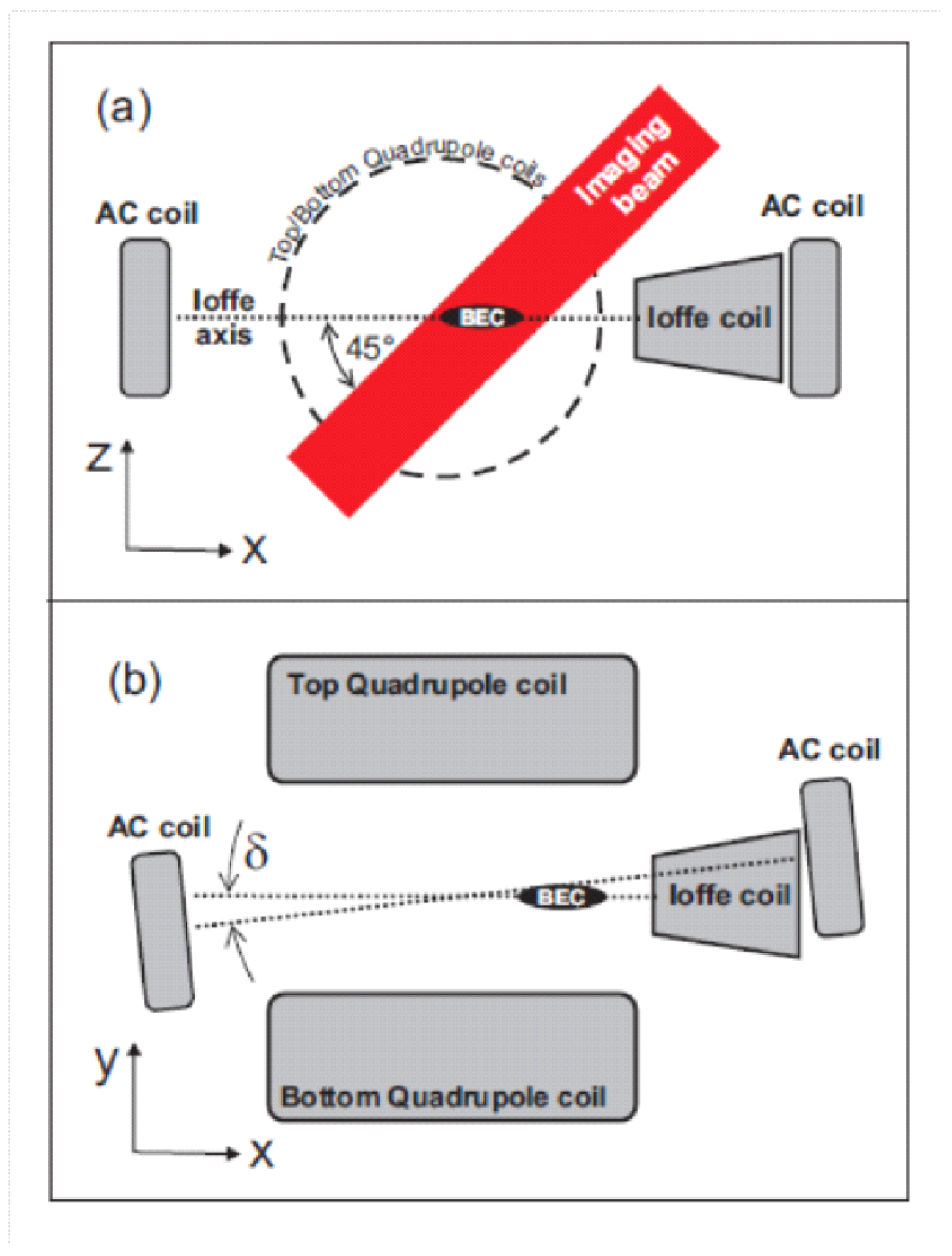}
\caption{Scheme of the main components composing the magnetic trap and the
additional extra coils producing the necessary oscillatory fields generating
trapped quantum turbulence.}\label{fig:FIG1}
\end{figure}

\begin{figure}[!h]
\centering
\includegraphics[scale=0.3]{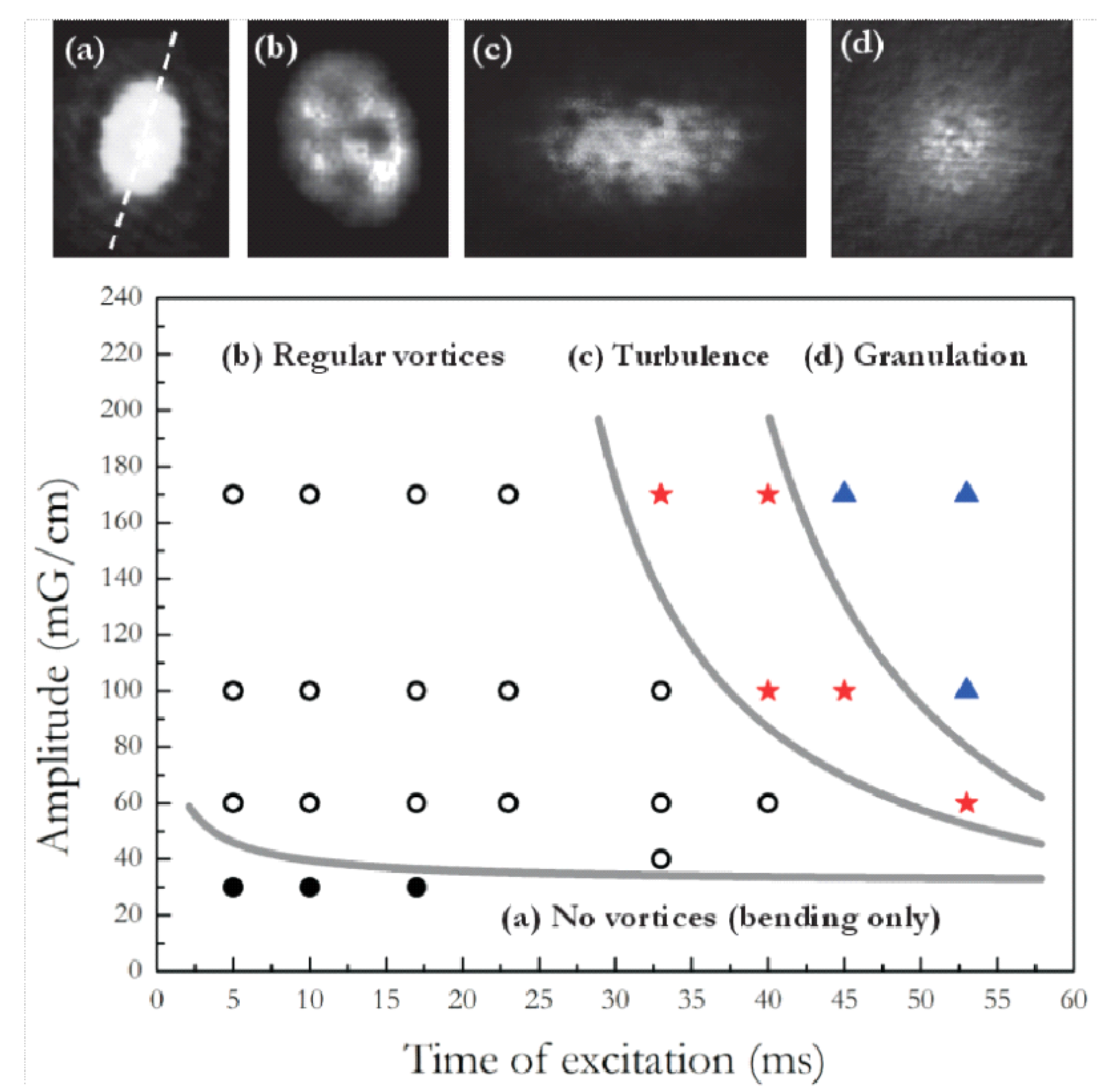}
\caption{Diagram, on the excitation amplitude-time plane, showing the observed
characteristic regions of four different identifiable structures, from the
simple bending of the superfluid axis, to the creation of vortices, vortex
tangle, and to granulation.}\label{fig:FIG2}
\end{figure}

In brief, we start with a BEC of Rb atoms confined in a harmonic trapping
potential with the frequencies $\omega_x=\omega_y=9\,\omega_z$ and 
$\omega_z = 2\pi\,\times 23 \,Hz$. The typically produced BEC contains 
$2 \times 10^5$ atoms. A pair of coils (as in Fig. \ref{fig:FIG1}) forms a 
magnetic field that mechanically excites the trapped condensate. The notation 
of axes in this figure corresponds to that used in the text with the interchange 
of $x$ and $z$. The excitation is achieved by applying an oscillatory current 
in the extra coils. The produced distortions of the trapping potential cause 
a combination of translations and rotations of the cloud. The result of such 
an excitation is a combination of the effects, going from a simple bending
of the symmetry axis of the cloud up to the generation of vortices in many
directions with a final granulation of the cloud. We have characterized the
overall behavior of the system in a diagram presenting the regions of
observations in Fig. \ref{fig:FIG2} \cite{Seman_109}.

Small amplitudes of oscillation can only produce a bending mode intrinsically
connected to the scissor mode \cite{Marago_84} present in atomic trapped
superfluids. Larger amplitudes of oscillation, combined with longer excitation
times, can produce vortices with a characteristic array of QT. As is shown in
Fig. \ref{fig:FIG2}, in a range of the excitation parameters, there arise
vortices directed along the cloud axis, but still not yet showing a fully
tangled configuration. Quantum turbulence takes place in the region of
parameters with a clear separation of behavior between the regions.

The generation of vortices takes place because the oscillation of the atomic
superfluid cloud produces collective modes \cite{Seman_107} leading to the 
generation of coherent modes \cite{Yukalov_35}. In a more recent observation, 
it has been verified \cite{Schiavinatti_Tesis} that the excitation, through 
a combination of oscillations produces, together with collective modes,
the excitation of the second sound mode coupled to the dipole mode. This
excitation corresponds to the counterflow between condensate and thermal cloud,
with the possible generation of vortices in the regions of the maximal relative
motion. At low temperatures, when the normal fraction is negligible, dynamic
instability appears due to the counterflow between different parts of the
condensate \cite{Yukalov_76,Dutton_77,Yukalov_78,Ruostekoski_79,Shomroni_80,Ma_81,
Ishiro_82,Simula_83}.

Being generated, vortices can be distributed in many directions, first, without
actually forming a tangled configuration. When the finite size of the cloud is
saturated with the vortices, any further pumped energy forces a fast evolution
of the vortices, promoting their reactions by reconnections \cite{Tsubota_121},
eventually yielding a turbulent cloud. At this stage, not only the distribution
of the vortices is an indication for the occurrence of turbulence, but also a
change in the hydrodynamic behavior, during the free expansion of the cloud,
works as the indicator of turbulence. While a non-turbulent cloud of an atomic
superfluid demonstrates the inversion of aspect ratio during free expansion,
the turbulent cloud preserves the original aspect ratio
\cite{Henn_106,Caracanhas_166}.

It has been observed that the existence of a boundary, between the regular and
turbulent superfluids, on the amplitude-time diagram of Fig. \ref{fig:FIG2}
is the consequence of the finite size of the cloud, as explained in 
Ref. \cite{Shiozaki_108}.

\begin{figure}[!h]
\centering
\includegraphics[scale=0.25]{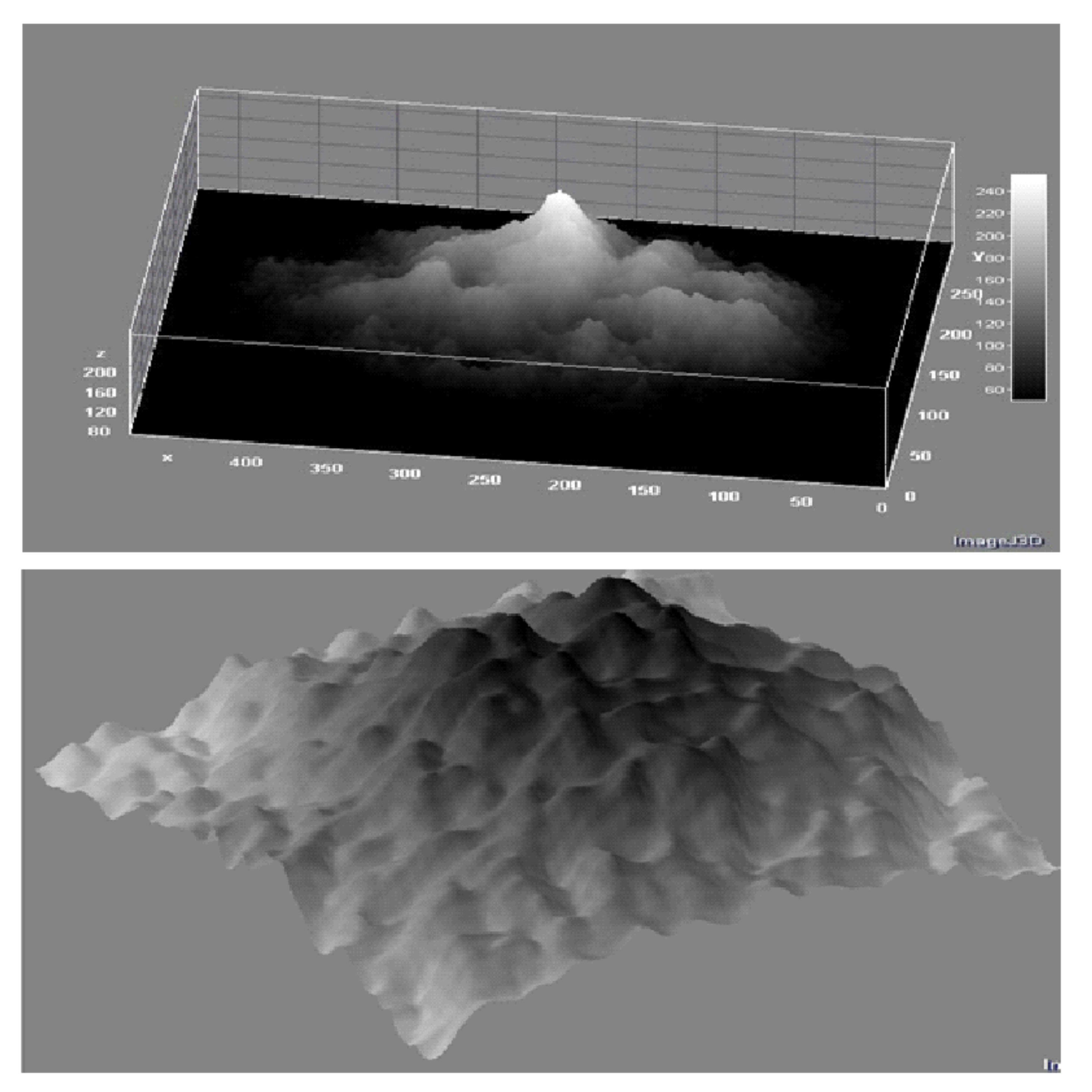}
\caption{The difference between a turbulent and a granulated cloud. While in 
the first case the landscape is made of vortex filaments distributed in space, 
the granulated state corresponds to a collection of grains characterizing 
strong density fluctuations.}\label{fig:FIG3}
\end{figure}

For the extreme case of excitation (high amplitude and longer excitation times),
the turbulent condensate evolves into a granulated state, when the original
condensate cloud breaks into many grains. The transition from the turbulent to
fragmented cloud is presented by the density profile of Fig. \ref{fig:FIG3}.

The experimental observation of the atomic cloud inside the trap is not easy. 
This is because the produced condensate has the size of just a few microns. 
To perform an absorption measurement in situ, we would be severely limited by 
diffraction. We therefore, first, allow a free expansion of the cloud, and then 
perform absorption measurements. For the observed states, discussed above, the 
time of flight before absorption was of 15 ms. In this case, the size of the 
cloud is many times larger than the actual size in situ. As far as, during the 
time of flight, the density is greatly reduced, the interactions are also 
reduced, freezing the existent structure, that now evolves much slower in time. 
It is a general consensus that the free expansion preserves the in situ 
structure of the atomic cloud. 

Figure 4 demonstrates the absorption image of the granulated cloud and the 
details showing the domains of the grains after free expansion of 15 ms. We 
observe an isotropic expansion and the details of the figure allow us to 
identify the grains arising in the originally homogeneous superfluid.  
Applying the reversibility analysis, we find that, in situ, the grains have 
the average size of about 0.25 microns. They are clearly not regular either 
in shape or in size and do not form any structure that could be observed 
through the absorption.

\begin{figure}[!h]
\centering
\includegraphics[scale=0.35]{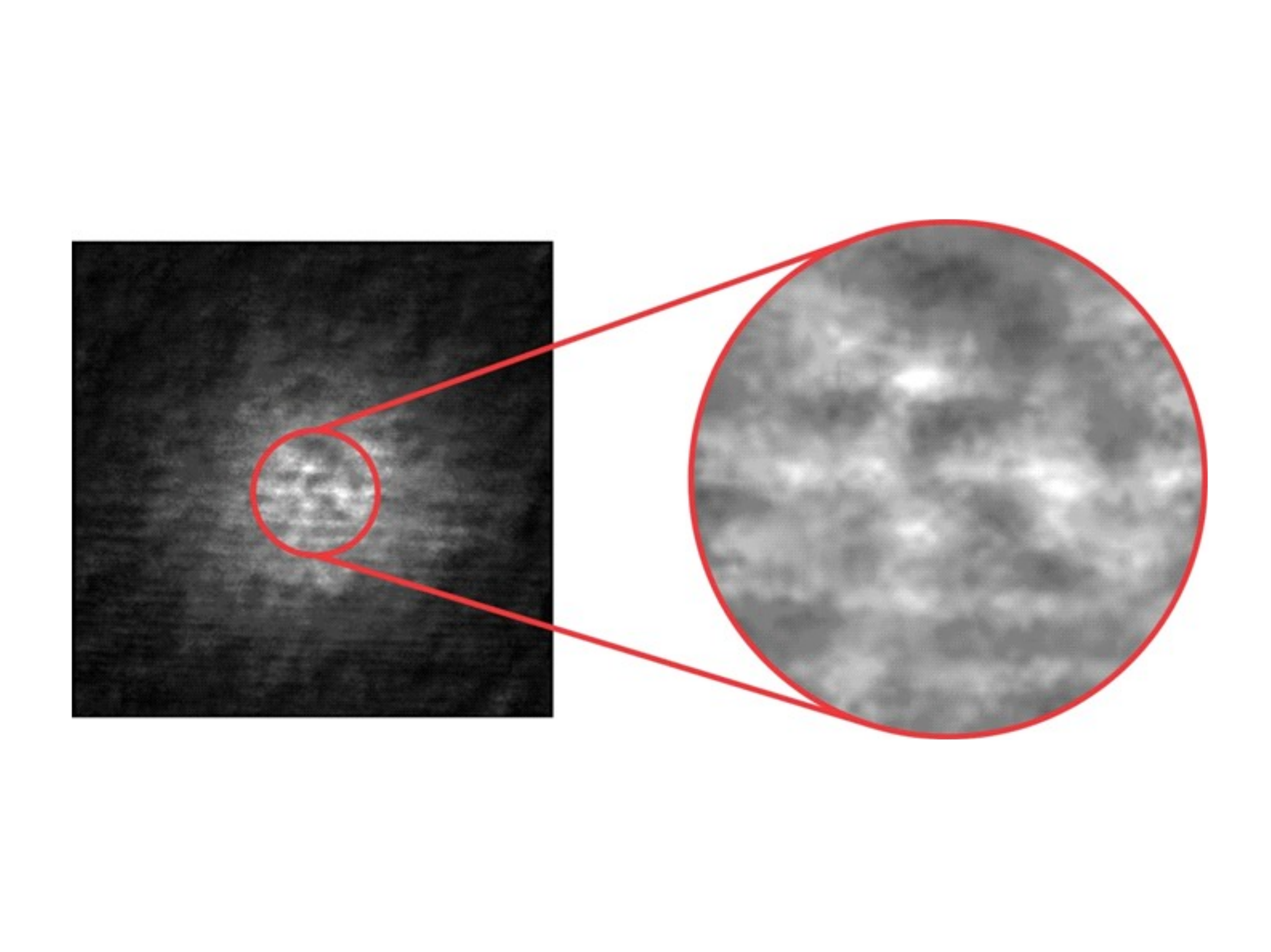}
\caption{Absorption by the granulated cloud showing the domains of the grains}
\label{fig:FIG4}
\end{figure}

For convenience, let us summarize the data characteristic of our experiments 
with $^{87}$Rb atoms. The mass of a Rb atom is $m = 1.445 \times 10^{-22}$ g.
The scattering length is $a_s = 0.577 \times 10^{-6}$ cm. The radial frequency
is $\omega_{\perp} = 1.32 \times 10^3$ s$^{-1}$ and longitudinal frequency is
$\omega_z = 1.445 \times 10^2$ s$^{-1}$. The corresponding oscillator lengths 
are $l_\perp = 0.744 \times 10^{-4}$ cm and $l_z = 2.248 \times 10^{-4}$ cm. 
The average oscillator frequency and length are 
$\omega_0 = 0.631 \times 10^3$ s$^{-1}$ and $l_0 = 1.076 \times 10^{-4}$ cm.
The effective condensate volume is $V_{eff} = 0.783 \times 10^{-11}$ cm$^3$.
The average condensate density is 
$\rho \sim 2.554 \times 10^{15}$ cm$^{-3}$. The mean interatomic 
distance is $a = 0.732 \times 10^{-5}$ cm, which is much larger than the 
scattering length. Hence, the gas is rarified. The gas parameters are small, 
$\rho a_s^3 = 0.491 \times 10^{-3}$ and $\rho ^{1/3} a_s = 0.079$. This implies
that atomic interactions are weak. However, the effective coupling 
parameter (44) is large, $g = 1.95 \times 10^4$. Therefore the corresponding
nonlinearity is very large.  

The trap is subject to an external field modulation during the time 
$t_{ext} = 0.02 - 0.06$ s, with an alternating potential of frequency
$\omega = 1.257 \times 10^3$ s$^{-1}$. The related modulation period is
$t_{mod} \equiv 2\pi / \omega = 5 \times 10^{-3}$ s. The local equilibration 
time is $t_{loc} = m /(\hbar \rho a_s) = 0.929 \times 10^{-4}$ s. This is
much shorter than the modulation period, because of which the system is always
in local equilibrium.
 
With the average grain size $l_g \sim 2.5 \times 10^{-5}$ cm, the number of 
atoms inside a grain is $(l_g/a)^3 \sim 40$. And the number of grains in the 
trap is of order $(l_0/l_g)^3 \sim 400$. 

The majority of experimental observations can be explained by the models
of Secs. 4 to 7. But, certainly, other experiments for cross-checking the
measured and theoretical dependencies are needed. Recent measurements of the
momentum distribution of a turbulent cloud show the existence of a power-law
type dependence $n (k)\propto k^{-\delta}$, which also requires confirmation
and analysis with respect to its relation to the Kolmogorov-type behavior.

\section{Conclusion}

We have described the procedure of exciting a system of trapped
Bose-condensed atoms by an external alternating potential, forcing the
system to pass through several qualitatively different stages. Initially,
the system is almost completely condensed, which is characterized by the
broken global gauge symmetry. Applying sufficiently strong external
perturbation transfers the system into a nonequilibrium state. First, there
appear separate vortices and antivortices, which marks the transfer from
the regular superfluid to vortex superfluid. Increasing perturbations is
realized by either strengthening the amplitude of the applied alternating
field or by its longer action on the system. Sufficiently strong
perturbation generates a variety of coherent topological modes that decay
into basic vortices with vorticity one. Thus, a multiplicity of vortices
and antivortices is effectively generated. The location of these vortices
inside the trap and their directions are random, which is caused by the
absence of a unique rotation axis of the applied alternating potential.
Because of this, the increasing perturbation creates not a vortex lattice,
as it would be in the case of a uniaxial rotation, but forms en ensemble
of randomly directed vortices. When the number of vortices becomes large,
they form a random vortex tangle typical of quantum turbulence. Increasing
further the amount of energy, injected into the trap, breaks the system
into pieces. Then Bose-condensed  grains, or droplets, are surrounded by
uncondensed gas in the normal state. Pumping more energy into the trap
reduces the fraction of condensed atoms. Finally, the system should transfer
into the normal state, where the global gauge symmetry is restored.

Thus, starting with a regular superfluid, we pass through the states of
vortex superfluid, turbulent superfluid, granular fluid, and should finish
with chaotic fluid that is in a state of classical turbulence. In that way,
the initial state with global gauge symmetry breaking is transformed,
through a sequence of qualitatively different regimes, to a state with
the restored global gauge symmetry. Transitions between different regimes
are classified as crossovers, though they are sufficiently sharp for
allowing us to define the corresponding crossover lines. All these
transitions, except that to chaotic fluid, are illustrated by experiments
with trapped $^{87}$Rb atoms.

\vskip 2cm

{\bf Acknowledgement}

We are grateful to all our co-authors for collaboration. One of the authors
(V.I.Y.) acknowledges financial support from the Russian Foundation for
Basic Research.

\end{document}